\documentstyle[sprocl]{article}

\input{psfig}

\def\be{\begin{equation}}
\def\ee{\end{equation}}
\def\bea{\begin{eqnarray}}
\def\eea{\end{eqnarray}}

\begin{document}

\title{FINE-TUNING CARBON-BASED LIFE IN THE UNIVERSE 
BY THE TRIPLE-ALPHA PROCESS IN RED GIANTS}

\author{HEINZ OBERHUMMER}
\address{Institute of Nuclear Physics, Vienna University of 
Technology, Wiedner Haupstr.\ 8-10, A-1040 Wien, Austria \\
E-mail: ohu@kph.tuwien.ac.at}
\author{ATTILA CS\'OT\'O}
\address{Department of Atomic Physics, E\"otv\"os University,
P\'azm\'any P\'eter s\'et\'any 1/A, H-1117 Budapest, Hungary \\ 
E-mail: csoto@nova.elte.hu} 
\author{HELMUT SCHLATTL}
\address{Max-Planck Institut f\"ur Astrophysik,
Karl-Schwarzschild-Str.~1, D-85740 Garching, Germany \\
E-mail: schlattl@MPA-Garching.MPG.DE} 


\maketitle\abstracts{Through the triple-alpha process occurring
in red giant stars the bulk of the carbon existing in our universe
is produced. We calculated the change of the triple-alpha
reaction rate for slight variations of the nucleon-nucleon
force using a microscopic 12-body model. Stellar model
calculations for a low-mass, intermediate-mass and massive
star using the different triple-alpha reaction rates
obtained with different strengths of the N-N interaction
have been performed. Even with a change of 0.4\% in the strength of 
N-N force, carbon-based
life appears to be impossible, since all the stars then would
produce either almost solely carbon or oxygen, but could not 
produce both elements.
}

\section{Introduction}
The Anthropic Principle in its weak form can be formulated in the following way:
{\it The observed values of all physical and cosmological quantities are not equally
probable but they take on values restricted by the requirement
that there exist sites where carbon-based life can evolve and by the requirement
that the universe be old enough for it to have already done so\/} \cite{BA86}.
The triple-alpha process occurring in the helium burning of red giants
is of special significance with respect to the anthropic principle.
Through this process the bulk of carbon, e.g.,
the carbon isotope $^{12}$C, existing in our universe
is produced.  Carbon is synthesized further by alpha
capture to oxygen, e.g., to the oxygen isotope $^{16}$O, leading
to abundance ratio of $^{12}$C :  $^{16}$O $\approx$ 1 : 2.
Without the production of an appreciable
amount of carbon, obviously
no carbon-based life could have developed in the universe.
On the other hand, the production of oxygen is also absolutely necessary, 
because no spontaneous development of carbon-based life
is possible without the existence of water (H$_2$O).

The reason for the relevance
of the triple-alpha process with respect to the anthropic principle
lies also in the fact that one has to deal with physical quantities that
are in the realm of experimentally verifiable and theoretically calculable
physics. This is for instance hardly the case for the much less well-known and
complicated science necessary for the description of the Big Bang as well as for the 
creation and evolution of life on earth.

The formation of $^{12}$C in hydrogen burning
is blocked by the absence of stable elements at 
mass numbers $A=5$ and $A=8$. \"Opik and Salpeter \cite{OP51,SA52} pointed out
that the lifetime of $^8$Be is long enough,
so that the $\alpha +\alpha \rightleftharpoons\; $$^8$Be
reaction can produce macroscopic amounts of equilibrium
$^8$Be in stars. Then, the unstable $^8$Be could capture an
additional $\alpha$-particle to produce stable $^{12}$C.
However, this so-called triple-alpha reaction has very
low rate since the density of $^8$Be in the stellar
plasma is very low, because of its short lifetime of 10$^{-16}$\,s.

Hoyle \cite{HO53} argued  that in order to explain the
measured abundance of carbon in the Universe, the triple-alpha
reaction cannot produce enough carbon in a non-resonant way, therefore 
it must proceed through a hypothetical resonance of
$^{12}$C, thus strongly enhancing the cross section.
Hoyle suggested that this resonance is a $J^\pi=0^+$ state
at about $\varepsilon=0.4$ MeV (throughout this paper $\varepsilon$ denotes
resonance energy in the center-of-mass frame relative to
the three-alpha threshold, while $\Gamma$ denotes the full
width). Subsequent experiments indeed found a $0^+$
resonance in $^{12}$C in the predicted energy region by
studies of the reaction $^{14}$N(d,$\alpha$)$^{12}$C
and the $\beta^-$-decay of $^{12}$B~\cite{HO53,CO57}. It
is the second $0^+$ state ($0^+_2$) in $^{12}$C. Its modern 
parameters~\cite{Ajzenberg}, $\varepsilon=0.3796$\,MeV and 
$\Gamma=8.5 \times 10^{-6}$\,MeV, agree well with the old 
theoretical prediction.

In this work we investigate the amount of carbon
and oxygen production in helium burning of red giants by slightly varying the
nucleon-nucleon interaction. In Ref.~\cite{LI89}
only hypothetical {\it ad hoc\/} shifts of the resonance energy
of the $0_2^+$-state were investigated, whereas in this work we start by
variations of the underlying N-N interaction.

After the Introduction we discuss in Sect.~2 our microscopic
three-cluster model, the effective
nucleon-nucleon (N-N) interactions, and the complex scaling me\-thod, 
which is used to describe the $0^+_2$ state of $^{12}$C.
In Sect.~3 the change of the triple-alpha reaction rates
by slightly varying the underlying N-N interaction is calculated.
In Sect.~4 we present the results for stellar model
calculations for a low-mass, intermediate-mass and massive
star using the triple-alpha reaction rates
obtained with different
strengths
of the effective N-N in\-ter\-ac\-tion. In Sect.~5 the
obtained results are summarized.

\section{Nuclear physics}
The astrophysical models that determine the amount of carbon and oxygen 
produced in red giant stars need some nuclear properties of $^{12}$C as
input parameters. Namely, the position and width of the $0^+_2$
resonance, which almost solely determines the triple-alpha reaction 
rate, and the radiative decay width for the $0^+_2\rightarrow 2^+_1$ 
transition in $^{12}$C. Here we calculate these quantities in a
microscopic 12-body model. 

In order to make such a calculation feasible, we use the microscopic
cluster model. This approach assumes that the wave function of certain
nuclei, like $^{12}$C, contain, with large weight, components which
describe the given nucleus as a compound of 2-3 clusters. By assuming
rather simple structures for the cluster internal states, the relative
motions between the clusters, which are the most important degrees of
freedom, can be treated rigorously. The strong binding of the free
alpha-particle ($^4$He) makes it natural that the low-lying states of
$^{12}$C are largely $3\alpha$-structures \cite{3alpha}. Therefore, our
cluster-model wave function for $^{12}$C looks like
\begin{equation}
\label{wfn}
\Psi^{^{12}{\rm C}}=\sum_{l_1,l_2} {\cal A} \Bigl
\{ \Phi^\alpha \Phi^\alpha \Phi^\alpha\chi^{\alpha(
\alpha\alpha)}_{[l_1l_2]L} (\mbox{\boldmath$\rho$}_1,
\mbox{\boldmath$\rho$}_2) \Bigl\}.
\end{equation}
Here ${\cal A}$ is the intercluster antisymmetrizer, the $\Phi^\alpha$ 
cluster internal states are translationally invariant $0s$ 
harmonic-oscillator shell-model states with zero total spin, the 
$\mbox{\boldmath$\rho$}$ vectors are the intercluster Jacobi 
coordinates, $l_1$ and $l_2$ are the angular momenta of the two 
relative motions, $L$ is the total orbital angular momentum and 
$[\ldots]$ denotes angular momentum coupling. The total spin and 
parity of $^{12}$C are $J=L$ and $\pi=(-1)^{l_1+l_2}$, respectively.

We want to calculate the resonance energy of the $0^+_2$ state in
$^{12}$C, relative to the $3\alpha$-threshold, and the 
$0^+_2\rightarrow 2^+_1$ radiative (E2) width, while slightly varying the
strength of the effective nucleon-nucleon (N-N) interaction. This way, the
sensitivity of the carbon and oxygen production on the effective
N-N interaction
can be studied. The $0^+_2$ state is situated above the
$3\alpha$-threshold, therefore for a rigorous description one has to
use an approach which can describe three-body resonances correctly. We
choose the complex scaling method \cite{CSM} that has already been used
in a variety of other nuclear physics problems, see e.g.~\cite{CSMappl}. 
In this method the eigenvalue problem of a transformed Hamiltonian
\begin{equation}
\label{cs}
\widehat{H}_\theta=\widehat{U}(\theta)\widehat{H}
\widehat{U}^{-1}(\theta)
\end{equation}
is solved, instead of the original $\widehat{H}$, where the
transformation $\widehat{U}$ acts on a function $f({\bf r})$ as 
\begin{equation}
\widehat{U}(\theta)f({\bf r})=e^{3 i \theta /2}f({\bf r}e^{i\theta}).
\end{equation}
Resonances are eigenvalues of $\widehat{H}$ with  
$E_{\rm res}=\varepsilon-i\Gamma/2$ ($\varepsilon,\, 
\Gamma >0$) complex energy, where $\varepsilon$ is the position of the
resonance while $\Gamma$ is the full width. A straightforward description
of such states are difficult because their wave functions are
exponentially divergent in the asymptotic region. The effect of the
Eq.~(\ref{cs}) complex scaling transformation is that the
positive-energy continuum of $\widehat{H}$ gets rotated (by the angle
$2\theta$) down into the complex energy plane, while the wave
function of any resonance becomes square-integrable if 
$2\theta>|{\rm arg}\, E_{\rm res}|$. Using this 
method \cite{3alpha}, we were able to localize the energy of the 
$0^+_2$ state of $^{12}$C.

The other important quantity that needs to be calculated is the
radiative width of the $0^+_2$ state, coming from the electric dipole
(E2) decay into the $2^+_1$ state of $^{12}$C. This calculation involves
the evaluation of the E2 operator between the initial $0^+_2$ 
three-body scattering state and the final $2^+_1$ bound state \cite{Gamma}. 
The proper three-body scattering-state treatment of the $0^+_2$
initial state is not feasible for the time being, therefore we use a
bound-state approximation to it. This is an excellent approximation for
the calculation of $\Gamma_\gamma$ because the total width of the
$0^+_2$ state is very small (8.5 eV \cite{Ajzenberg}). The value of 
$\Gamma_\gamma$ is rather sensitive to the energy difference between the
$0^+_2$ and $2^+_1$ states, so we have to make sure that the experimental
energy difference is correctly reproduced.

In order to see the dependence of the results on the chosen effective
N-N interaction, we performed the calculations using three different
forces. The Minnesota (MN) \cite{MN}, and the rather different Volkov 1
and 2 (V1, V2) \cite{Volkov} forces achieve similar quality in
describing light nuclear systems. We slightly adjusted a parameter (the
exchange mixture) of each forces in order to get the $0^+_2$ resonance
energy right at the experimental value. This leads to $u=0.941$ for MN
and $m=0.568$ and $m=0.594$ for V1 and V2, respectively. The results
coming from these forces are our baseline predictions. Then, we
multiplied the strengths of the forces by a factor $p$, which took up
the values $p=0.996$, 0.998, 0.999, 1.001, 1.002, and 1.004,
respectively, and calculated the resonance energies and gamma widths
again. This way we can monitor the sensitivity of the $^{12}$C
properties, and thus the carbon and oxygen production in stellar
environments, as the function of the N-N interaction strength. 
The results for the resonance energies and widths are shown in 
Table \ref{tab1}.
\begin{table}[t]  
\caption{The energy, $E_r$ (in keV), and gamma width, $\Gamma_{\gamma}$ 
(in meV), of the $0^+_2$-resonance as a function of the factor $p$}
\vspace{0.2cm}
\begin{center}
\begin{tabular}{ccccccc}
\hline
\multicolumn{1}{c}{N-N interaction}&
\multicolumn{2}{c}{MN}&
\multicolumn{2}{c}{V1}&
\multicolumn{2}{c}{V2}\\
\hline
\multicolumn{1}{c}{$p$}&
\multicolumn{1}{c}{$\varepsilon(p)$}&
\multicolumn{1}{c}{$\Gamma_{\gamma}(p)$}&
\multicolumn{1}{c}{$\varepsilon(p)$}&
\multicolumn{1}{c}{$\Gamma_{\gamma}(p)$}&
\multicolumn{1}{c}{$\varepsilon(p)$}&
\multicolumn{1}{c}{$\Gamma_{\gamma}(p)$}\\
\hline
1.004 & 273.3  & 2.85 & 294.4 & 3.21 & 306.0 & 3.28\\
1.002 & 327.5  & 2.91 & 337.5 & 3.26 & 343.7 & 3.31\\
1.001 & 353.7  & 2.93 & 358.7 & 3.27 & 361.7 & 3.32\\
1.000 & 379.6  & 2.96 & 379.6 & 3.29 & 379.6 & 3.33\\
0.999 & 405.2  & 2.99 & 400.3 & 3.31 & 397.2 & 3.34\\
0.998 & 430.5  & 3.01 & 420.8 & 3.32 & 414.6 & 3.34\\
0.996 & 481.4  & 3.07 & 460.7 & 3.34 & 450.0 & 3.34\\
\hline
\end{tabular}
\end{center}
\label{tab1}
\end{table}

As we mentioned, we correctly reproduced the $0^+_2-2^+_1$ energy
difference for $p=1.0$. This required the use of a slightly different
force for $2^+_1$. The resulting $\Gamma_\gamma$ values should be
compared to the experimental figure, $3.7\pm0.5$ meV \cite{Ajzenberg}.
Our model performs well, considering the fact that no effective charges
were used in the calculations. Using an effective charge $e_{\rm eff}$
for both the protons and neutrons would lead to $\Gamma_\gamma$
multiplied by $(1+2e_{\rm eff})^2$. 

\section{Triple-alpha reaction rate}
The results shown in Table \ref{tab1} can be used to calculate the
triple-alpha reaction rates.
The reaction rate for the triple--alpha process proceeding via the ground
state of
$^8$Be and the $0^+_2$--resonance in $^{12}$C is given by \cite{RO88}
\begin{equation}
\label{3alpha}
r_{3\alpha} = 3^{\frac{3}{2}} N_{\alpha}^3
\left(\frac{2 \pi \hbar^2}{M_{\alpha} k_{\rm B} T}\right)^3
\frac{\omega \gamma}{\hbar} \exp \left(-\frac{\varepsilon}{k_{\rm B} T}\right),
\end{equation}
where $M_{\alpha}$ and $N_{\alpha}$ is the mass and the number density of the
$\alpha$--particle, respectively. The temperature of the stellar plasma
is given by $T$. The quantity $\varepsilon$ denotes the difference in energy
between
the  $0^+_2$--resonance in $^{12}$C and the 3$\alpha$--particle
threshold. The resonance strength $\omega \gamma$ is given by
\begin{equation}
\label{omega}
\omega \gamma = \frac{\Gamma_{\alpha} \Gamma_{\rm rad}}
{\Gamma_{\alpha} + \Gamma_{\rm rad}} \approx \Gamma_{\gamma} .
\end{equation}

The approximation of the above expression for the decay widths of the
$0^+_2$--resonance
follows, because for the $\alpha$--width $\Gamma_{\alpha}$, the 
radiation width $\Gamma_{\rm rad}$, the electromagnetic decay width
$\Gamma_{\gamma}$
to the first excited state of $^{12}$C,
and the electron--positron pair emission
decay width $\Gamma_{\rm pair}$ into the ground state of $^{12}$C
the following approximations hold:
(i)  $\Gamma_{\alpha} \gg \Gamma_{\rm rad}$ and
(ii) $\Gamma_{\rm rad} = \Gamma_{\gamma} + \Gamma_{\rm pair} \approx
\Gamma_{\gamma}$.
Therefore, Eq.~(\ref{3alpha}) can be approximated by:
\begin{equation}
\label{alphaa}
r_{3\alpha} \approx 3^{\frac{3}{2}} N_{\alpha}^3
\left(\frac{2 \pi \hbar^2}{M_{\alpha} k_{\rm B} T}\right)^3
\frac{\Gamma_{\gamma}}{\hbar} \exp \left(- \frac{\varepsilon}{k_{\rm B} T}
\right) ,
\end{equation}

The two quantities in Eq.~(\ref{alphaa}) that change their value by varying the effective
N-N interaction are the energy of
the $0^+_2$--resonance $\varepsilon$ in $^{12}$C
and its electromagnetic decay width
$\Gamma_{\gamma}$. However,
the change in the reaction rate by varying the effective
N-N interaction with
$\varepsilon$ in the exponential factor of Eq.~(\ref{alphaa})
is much larger than the linear change with $\Gamma_{\gamma}$.

\section{Stellar burning}
The significance of low-mass (with masses less than about 2\,$M_{\odot}$,
intermediate-mass (between about $2M_{\odot}$ 
and $10M_{\odot}$), and massive stars (with masses more than about 
$10M_{\odot}$) in the nucleosynthesis of carbon is still not quite 
clear \cite{GU98}. Some authors claim that low-mass and 
intermediate-mass stars dominate in the production of carbon, 
whereas others favor the production of carbon in massive stars. In a
recent investigation \cite{GU98} by spectral analysis of solar-type 
stars in the Galactic Disk, the results are
consistent with carbon production in massive stars but inconsistent
with a main origin of carbon in low-mass stars. The significance
of intermediate-mass stars for the production of carbon in the
Galaxy is still somewhat unclear. Therefore we performed a
stellar model calculation for a typical low-mass, intermediate-mass,  
and massive star, respectively.

The calculation of the stellar models are performed with a contemporary
stellar evolution code, which contains the latest physics input.
In particular, using this code, up-to-date solar
models \cite{SW99} can be produced as well as 
the evolution of low-mass stars can be followed through the thermal
pulses of asymptotic giant branch-(AGB) stars \cite{WaWe}. The
nuclear network is designed 
preferentially to calculate the hydrogen and helium burning phases in 
low-mass stars. Additionally, the basic reactions of carbon and oxygen
burning are included, which may destroy the produced carbon and oxygen
in massive stars.

In this work the evolution of a 1.3, 5, and 20 solar
mass star is  calculated, which should represent the typical
evolution of low-mass, in\-ter\-me\-di\-ate-mass, and massive stars, 
respectively. The stars are followed from the onset of
hydrogen burning until the first thermal pulses in the AGB,
or until the core temperature reaches $10^9$\,K in the case of the 
$20M_{\odot}$ star, as the nuclear network is not sufficient to go
beyond this phase.

Large portions of the initial mass of a star are returned to the
interstellar medium (ISM) through stellar winds basically during the thermal 
pulse phase. Unfortunately, basically due to the simple convection model
used in stellar modeling, the composition of the wind can
not yet be determined very accurately from stellar evolution theory. 
However, it is beyond the scope of the present investigations to
determine how and when the material is return back to the ISM.
Instead we investigate how much C and O is produced altogether in
the star, which then may be blown away by stellar winds.

\begin{figure}[tb]
\begin{center}
\leavevmode
\psfig{figure=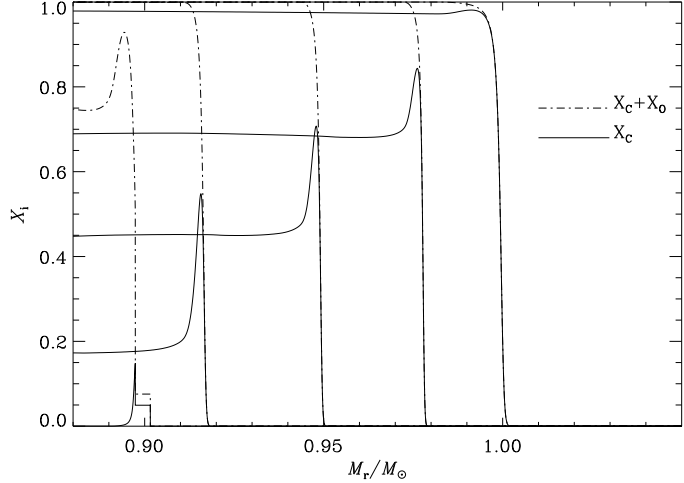,height=7cm}
\end{center}
\caption[]{The radial abundance profile of carbon and oxygen within a
5.0\,$M_\odot$ shortly before the first thermal pulse. The quantities 
$X_{\rm C}$ and $X_{\rm O}$ denote the radial abundances of
carbon and oxygen, respectively. The models are
from left to right with $p=0.996$, 0.999, 1.00, 1.001 and 1.004.\hfill
\label{Fig1}}
\end{figure}

For the three masses quoted above, the evolution of stars is 
calculated using the predictions of the MN interaction with 
$p=0.996$, 0.999, 1.000, 1.001, and 1.004. In Fig.~\ref{Fig1} the
radial abundance profiles of C and O are shown for a $5M\odot$ 
star, shortly before the first thermal pulse. With the standard cross 
section for the triple-alpha reaction (He-burning front at $M_{\rm
r}=0.95M_\odot$) roughly the same amount of carbon and oxygen is
produced. Changing the N-N interaction strength by only about 0.1\%, 
this ratio could already be altered significantly. If the N-N 
interaction strength is enhanced or reduced by 0.4\%, respectively,
then almost no oxygen or carbon is produced in a $5M_\odot$ star. 
Similar behavior can be observed for the $1.3M_\odot$ star, while in 
the case of a $20M_{\odot}$ star, the carbon can be destroyed almost 
totally even for $p=0.999$.

Thus, even with a minimal change of 0.4\% in the strength of the N-N force, 
carbon-based life appears to be impossible, since all the stars then 
would produce either almost solely carbon or oxygen, but could not  
produce both elements. For smaller variations in the N-N interaction 
the development of life depends on the stellar population, as for 
instance a $20M_\odot$ star with $p=0.999$ does not produce carbon, 
while the $1.3M_\odot$ and $5M_\odot$ stars still do. 

\section{Summary}
We have investigated the change of the carbon and oxygen production in
the helium burning of low-mass, intermediate-mass and massive stars by 
varying the underlying nucleon-nucleon interaction. 
A slight variation of the nucleon-nucleon interaction leads to a
change of the resonance energy of the second $0^+$ state ($0_2^+$) in 
$^{12}$C, thus drastically modifying the of the triple-alpha reaction 
rate. The changes in the relevant nuclear parameters in the
triple-alpha reaction rate, e.g.\ the resonance energy and radiative
decay width of the $0_2^+$ state in $^{12}$C, were obtained with
the help of a microscopic three-cluster model.

The impact of changing the triple-alpha reaction
rate by varying the effective N-N interaction in
the evolution of 1.3, 5 and 20 solar
mass stars is  calculated by following the typical
evolution of a low-mass, intermediate-mass, and massive star,
respectively. The calculations were carried out with a contemporary
stellar evolution code, which contains the latest physics input.
The result is that even with a small change of 0.4\% in the N-N
interaction strength, carbon-based
life appears to be impossible, since all the stars then would
produce either almost solely carbon or oxygen, but could not produce
both elements.

\section*{Acknowledgments}
This work was supported by OTKA Grants F019701, F033044, and 
D32513 (Hungary), and by the Bolyai Fellowship of the Hungarian 
Academy of Sciences. We also acknowledge the support by the Fonds
zur wissenschaftlichen Forschung in \"Osterreich, project P10361-PHY.

\end{document}